\documentclass[prb,twocolumn,amssymb,floatfix,showpacs,showkeywords,amsmath]{revtex4}
\usepackage{amsfonts}
\usepackage{bm}
\usepackage{graphicx}

\newcommand{\bS}{{\bm S}}
\newcommand{\bT}{{\bm T}}
\newcommand{\hT}{{\hat T}}
\newcommand{\br}{{\bm r}}
\newcommand{\bQ}{{\bm Q}}
\newcommand{\bq}{{\bm q}}

\newcommand{\hn}{\hat{n}}

\newcommand{\Ham}{{\mathcal{H}}}

\newcommand{\llangle}{\langle\kern-.25em\langle}
\newcommand{\rrangle}{\rangle\kern-.25em\rangle}

\begin{document}

\title{Spin-orbital frustrations and anomalous metallic state in iron-pnictide superconductors}

\author{Frank Kr\"uger$^1$}
\author{Sanjeev Kumar$^{2,3}$}
\author{Jan Zaanen$^2$}
\author{Jeroen van den Brink$^{2,4}$}

\affiliation{$^1$Department of Physics, University of Illinois, 1110 W. Green St., Urbana, IL 61801\\
$^2$Instituut-Lorentz, Universiteit Leiden, P. O. Box 9506, 2300 RA Leiden, The Netherlands\\
$^3$Faculty of Science and Technology, University of Twente, P. O. Box 217, 7500 AE Enschede, The Netherlands\\
$^4$Institute for Molecules and Materials, Radboud Universiteit Nijmegen, P. O. Box 9010, 6500 GL Nijmegen, The Netherlands
}

\begin{abstract}
We develop an understanding of the anomalous metal state of the parent compounds of recently discovered iron based superconductors starting from a strong coupling viewpoint, including orbital degrees of freedom. On the basis of an intermediate-spin ($S$=1) state for the Fe$^{2+}$ ions, we derive a Kugel-Khomskii spin-orbital Hamiltonian for  the active $t_{2g}$ orbitals. It turns out to be a highly complex model with frustrated spin and orbital interactions. We compute its classical phase diagrams and provide an understanding for the stability of the various phases by investigating its spin-only and orbital-only limits. The experimentally observed spin-stripe state is found to be stable over a wide regime of physical parameters and can be accompanied by three different types of orbital orders. Of these the orbital-ferro and orbital-stripe orders are particularly interesting since they break the in-plane lattice symmetry -- a robust feature of the undoped compounds. We compute the magnetic excitation spectra for the effective spin Hamiltonian, observing a strong reduction of the ordered moment, and point out that the proposed orbital ordering pattern can be measured in resonant X-ray diffraction. 
\end{abstract}

\date{\today}

\pacs{74.25.Jb, 74.70.-b, 74.20.Mn, 74.25.Ha}

\maketitle

\section{Introduction}

The begining of this year marked the discovery of a new and very unusual family of high temperature superconductors: the iron pnictides. Superconductivity at 26 K was discovered in fluorine doped rare-earth iron oxypnictide LaOFeAs \cite{Hosono1, Hosono2}. In subsequent experimental studies involving different rare earth elements a superconducting $T_c$ larger than 50 K was reported~\cite{Ren,Chen1,Chen2}. Since then a large number of experimental and theoretical papers have been published, making evident the immense interest of the condensed matter community in this subject~\cite{Norman}.

It has become clear that the iron pnictide superconductors have, besides a number of substantial differences, at least one striking similarity with the copper oxides: the superconductivity emerges by doping an antiferromagnetic, non-superconducting parent compound. This antiferromagetism is however of a very unusual kind. Instead of the simple staggered  '$(\pi,\pi)$' antiferromagnetism of the undoped cuprates, this 'stripe'  or '$(\pi,0)$' spin order involves rows of parallel spins on the square Fe-ion lattice that are mutually staggered~\cite{dai}. In fact, before this order sets in a structural phase transition occurs where the two in-plane lattice constants become inequivalent.  This structural distortion is very small, but it appears that the electron system undergoes a major reorganization at this transition.  This is manifested by resistivity anomalies, drastic changes in the Hall- and Seebeck coefficients, and so forth~\cite{0806.3878}. Although the magnetic- and structural distortion appear to be coincident in the 122 family~\cite{goldman,dai}, in the 1111 compounds they are clearly separated~\cite{dai}, and there it is obvious that the large scale changes in the electron system occur at the structural transition while barely anything is seen at the magnetic transition.

Given that the structural deformation is minute, this is an apparent paradox. Assuming that only the spins matter one could envisage that the spin ordering would lead to a drastic nesting type reorganization of the Fermi surfaces, causing a strong change in the electronic properties. But why is so little happening at the magnetic transition? One could speculate that the spins are fluctuating in fanciful ways, and that these fluctuations react strongly to the structural change~\cite{kivelson,sachdev,mazin}.  Such possibilities cannot be excluded on theoretical grounds but whichever way one wants to proceed invoking only spins and itinerant carriers: one is facing a problem of principle. 

This paper is dedicated to the cause that valuable lessons can be learned from the experiences with manganites when dealing with the pnictides. A crucial lesson learned over a decade ago, when dealing with the colossal magneto resistance (CMR) physics of the manganites, was the demonstration by Millis, Littlewood, and Shraiman~\cite{millis95} that the coupling between fluctuating spins and charge carriers can only cause relatively weak transport anomalies.  In the pnictides one finds that the resistivity drops by a couple of milliohm centimeters, that the Hall mobility increases by 2-3 orders of magnitude, and most significantly the Seebeck coefficient  drops by an order of magnitude from a high temperature limit order value of 40 $\mu$ V/K   in crossing the transition. It is very questionable if spin-carrier coupling of any kind, be it itinerant or strongly coupled, can explain such large changes in the transport properties.

\subsection{Role of Electron-Electron Interactions}

Comparing the pnictides with the cuprate superconductors there is now a consensus that in two regards these systems are clearly different: (i) in the pnictide system no Mott insulator has been identified indicating that they are 'less strongly correlated' than the cuprates in the sense of the Hubbard type local interactions; (ii) in the pnictide one has to account for the presence of several $3d$ orbitals playing a role in the low energy physics, contrasting with the single $3d_{x^2-y^2}$  orbital that is relevant in the cuprates. 

As a consequence, the prevailing viewpoint is to regard the pnictides as LDA metals, where the multi-orbital nature of the electronic structure gives rise to a multi-sheeted Fermi surface, while the 'correlation effects' are just perturbative corrections, causing moderate mass enhancements and so on. 

Although there is evidence that the system eventually discovers this 'Fermi-liquid fixed point'  at sufficiently low temperatures, it is hard to see how this can explain the properties of the metallic state at higher temperatures. The data alluded to in the above indicate pronounced 'bad metal' behavior, and these bad metal characteristics do not disappear with doping. In fact, one can argue that the term 'bad metal' actually refers to a state of ignorance: it implies that the electron system cannot possibly be a simple, coherent Fermi-liquid.

\subsection{Spin-Charge-Orbital Correlations}
Another important lesson from the manganites is that the presence of multiple orbitals can mean much more than just the presence of multiple LDA bands at the Fermi-energy.  Manganite {\em metals} have a degree of itineracy in common with the pnictides, but they still exhibit correlated electron physics tied to orbital degeneracy which is far beyond the reach of standard band structure theory.

The seminal work by Kugel and Khomskii in the 1970's made clear that in Mott insulators orbital degrees of freedom turn into dynamical spin like entities that are capable of spin-like ordering phenomena  under the condition that in the local limit one has a Jahn-Teller (orbital) degeneracy\cite{kugelkhomskii}. The resulting orbital degrees of freedom can have in dynamical regards a 'life of their own'. This manifests itself typically in transitions characterized by small changes in the lattice accompanied by drastic changes in the electronic properties. 

In the manganites there are numerous vivid examples of the workings of orbital ordering~\cite{Brink01,Brink99a,Brink99b}. Under the right circumstances one can find a transition from a high temperature cubic phase to a low temperature tetragonal phase accompanied by a quite moderate change in the lattice, but with a change in the electron system that is as drastic as a 'dimensional transmutation': this system changes from an isotropic 3D metal at high temperature to a quasi 2D electron system at low temperatures where the in-plane resistivity is orders of magntitude lower than the $c$-axis resistivity~\cite{Dho02,Akimoto98,Tokura97}.

The explanation is that one is dealing in the cubic manganite with a Mn$^{3+}$ ion with an $e_g$ Jahn-Teller degeneracy involving $3d_{x^2-y^2}$ and $3d_{3z^2-1}$ orbitals. In the low temperature 'A-phase' one finds a 'ferro' orbital order where cooperatively the $x^2$-$y^2$ orbitals are occupied. This greatly facilitates the hopping in the planes while for simple orthogonality reasons coherent transport along the $c$-axis is blocked.  Since the $d$-electrons only contribute modestly to the cohesive energy of the crystal, this large scale change in the low energy degrees of freedom of the electronic system reflect only barely in the properties of the lattice.  On the other hand, this orbital order is a necessary condition for the spin system to order, and at a lower temperature one finds a transition to a simple staggered antiferromagnet, in tune with the observation that in the A-phase the effective microscopic electronic structure is quite similar to the ones found in cuprate planes.

The ruthenates are another class of materials in which the orbital degrees of freedom play a decisive role, in both the metallic and insulating phases.  Bilayer Ca$_3$Ru$_2$O$_7$, for instance, has attracted considerable interest because the observed CMR-effect is possibly driven by orbital scattering processes among the conduction electrons~\cite{Lin05,Cuoco06}. Another example is Tl$_2$Ru$_2$O$_7$, in which below 120 K its 3D metallic state shows a dramatic dimensional reduction and freezes into a quasi-1D spin system, accompanied by a fundamental orbital reorganization~\cite{Lee06,Brink06}.

It is very remarkable that the groundstate of all iron pnictides is characterized by a very similar spatial anisotropy of the magnetic exchange interactions: along one direction in the plane the Fe-Fe bonds are strong and antiferromagnetic, whereas in the orthogonal direction they are very weak and possibly even ferromagnetic~\cite{savrasov}. With all the others, also this observation is consistent with our hypothesis that the undoped iron pnictides are controlled by 'spin-charge-orbital' physics, very similar in spirit to the ruthenates and manganites. 

\subsection{Organization of this Paper}
In Sec. II of this paper we derive the spin-orbital Hamiltonian starting with a three-orbital Hubbard model for the iron square lattice of the iron pnictides. The phase diagrams in the classical limit of this Hamiltonian are discussed in Sec. III. We analyze the various phase transitions by also considering the corresponding spin-only and orbital-only models. Sec. IV deals with the results on magnetic excitation spectra, which provide a possible explanation for the reduction of magnetic moment, a central puzzle in the iron superconductors. We conclude by commenting briefly on how the itineracy may go hand in hand with the orbital 'tweed' order that we put forward in the present study, and point out that the 'tweed' orbital ordered state can, in principle, be observed in resonant X-ray diffraction experiments.

\section{Spin-orbital model for Iron planes}
\label{sec.KK}

As stated above, the superconducting iron-pnictides are not strongly coupled doped Mott insulators. Staying within the realm of Hubbard-model language they are likely to be in the intermediate coupling regime where the Hubbard $U$'s are of order of the bandwidth. To at least develop qualitative insight in the underlying physics it is usually a  good idea to approach this regime from strong coupling for the simple reason that more is going on in strong coupling than in the weak coupling band structure limit. As the experience with for instance the manganites and ruthenates shows, this is even more true when we are dealing with the physics associated with orbital degeneracy. The orbital ordering phenomena that we have already alluded to, take place in itinerant systems but their logic is quite comprehensible starting from the strongly coupled side. 

Thus as a first step we will derive the spin-orbital model of pnictides starting from a localized electron framework. A condition for orbital phenomena to occur is then that the crystal fields conspire to stabilize an intermediate spin ($S=1$) ionic states.  These crystal fields come in two natural varieties: one associated with the tetrahedral coordination of Fe by the As atoms, and a tetragonal field associated with the fact that the overall crystal structure consists of layers. When these crystal fields would be both very large the Fe $3d^6$ ions would form a low spin singlet state. This is excluded by the observation of magnetism, and moreover band structure calculations indicate that the crystal fields are relatively small. 

The other extreme would be the total domination of Hund's rule couplings and this would result in a high spin $S=2$ state, which appears to be the outcome of spin polarized LDA and LDA+U calculations~\cite{giovannetti}. However, given that for elementary chemistry reasons one expects that the tetrahedral splitting is much larger than the tetragonal splitting there is the possibility that the Hunds rule overwhelms the latter but looses from the former, resulting in an  'intermediate' $S=1$ state. Although the issue is difficult to decide on microscopic grounds, for orbital physics to be relevant we need an intermediate spin state as in the present crystal field scheme this is the only ionic $d^6$ state that exhibits a Jahn-Teller groundstate degeneracy (see Fig. 1).

In this situation the starting Hubbard model involves a non-degenerate $|xy\rangle$ and two doubly-degenerate $|xz\rangle$ and  $|yz\rangle$ orbitals, as will be defined in subsection A. The details of the derivation of the model are given in subsection B. The derivation does not assume any specific structure for the hopping parameters and hence, is completely general. The algebra involved in the derivation is tedious but straightforward and a general reader may wish to skip subsection B and jump directly to subsection C where we discuss the relevant hopping processes for the Fe-As plane. Incorporating these hopping parameters leads to the model relevant to the iron plane.

\subsection{Hubbard model for pnictide planes for the intermediate-spin $d^6$ state }

\begin{figure}
\includegraphics[width=0.9\linewidth]{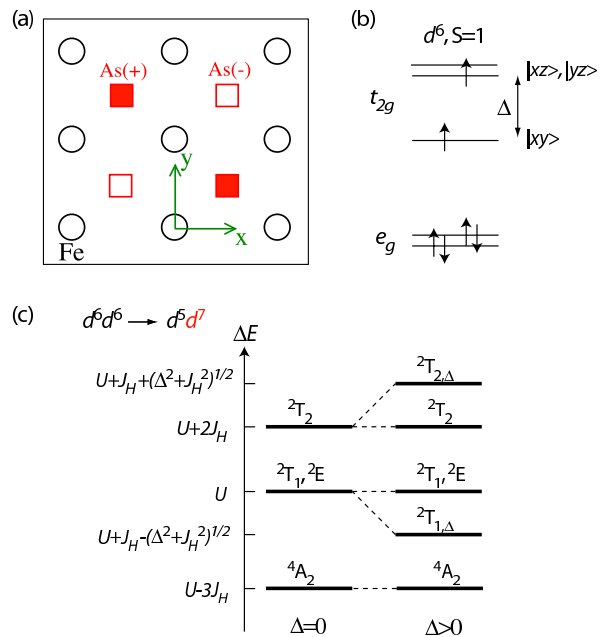}
\caption{(Color online) (a) Fe square lattice (black circles) and relative positions of the As ions. The latter are located in adjacent layers above (filled red squares) and below (empty red squares) the Fe plaquettes. (b) Schematic illustration of a ground-state $d^6$ configuration of the Fe ions corresponding to an intermediate $S=1$ spin state. (c) Multiplet structure of the $d_i^6d_j^6\rightleftharpoons d_i^7d_j^5$ charge excitations for localized $e_g$ electrons.}
\label{level}
\end{figure}

The iron ions are in a $d^6$ configuration where we assume the low lying $e_g$ orbitals to be fully occupied due to a large crystal-field splitting 
between the $e_g$ and $t_{2g}$ states. The two remaining electrons occupy the three $t_{2g}$ orbitals $|a\rangle:=|xz\rangle$,  $|b\rangle:=|yz\rangle$, and $|c\rangle:=|xy\rangle$ with $x$ and $y$ pointing along the bonds of the iron square lattice. Due to the Hund's coupling $J_H$ between the $t_{2g}$  electrons, such a configuration leads to an $S=1$ intermediate spin state of the $d^6$ Fe ions. Further, we incorporate a small tetragonal splitting $\Delta$ between the $|xy\rangle$ state and the $|xz\rangle$, $|yz\rangle$ doublet (see Fig. \ref{level}).

Assuming the $e_g$ electrons to be localized, the physical situation is very similar to almost cubic vanadates like YVO$_3$ or LaVO$_3$ where the two $d$-electrons of the V$^{3+}$ ions occupy nearly degenerate $t_{2g}$ orbitals. Interestingly, in theses systems orbital ordering in the presence of a small crystal-field splitting $\Delta$ can lead to C-type antiferromagnetism\cite{Oles+05,Oles+07,Horsch+08} characterized by an ordering wavevector $\bQ=(\pi,\pi,0)$. The effective Hubbard model for the $t_{2g}$ electrons consists of a kinetic energy part $\Ham_t$, a crystal field splitting $\Ham_\textrm{cf}$, and of the on-site electron-electron interactions $\Ham_\textrm{int}$,

\begin{equation}
\Ham=\Ham_t+\Ham_\textrm{cf}+\Ham_\textrm{int},
\end{equation}
with a kinetic energy contribution that is much richer than in the vanadates. For the nearest neighbor bonds the effective hoppings between the Fe $t_{2g}$ orbitals have contributions from both direct $d-d$ and $d-p-d$ processes via As $p$-orbitals. These As ions are located in adjacent layers above or below the Fe ion plaquettes as illustrated in Fig.~\ref{level}a. Because of this particular geometry, the indirect As mediated hoppings should be of similar strength for nearest and next-nearest neighbor Fe ions. At this point, we do not specify the effective hopping matrix elements $t^{(i,j)}_{\alpha,\beta}$ between orbitals $\alpha,\beta=a,b,c$ along a particular bond $(i,j)$ and write the kinetic energy operator in the most general form,

\begin{equation}
\Ham_t =- \sum_{(i,j)}\sum_{\alpha\beta,\sigma}t^{(i,j)}_{\alpha\beta}(d^\dagger_{i\alpha\sigma}d_{j\beta\sigma}+\textrm{h.c.}),
\label{Ht}
\end{equation}
where $d^\dagger_{i\alpha\sigma}$ ($d_{i\alpha\sigma}$) creates (annihilates) an electron on site $i$ in orbital $\alpha$ with spin $\sigma=\uparrow,\downarrow$. The crystal-field splitting between the $t_{2g}$ orbitals is simply given by

\begin{equation}
\Ham_\textrm{cf} = \sum_{i\alpha} \epsilon_\alpha \hn_{i\alpha},
\label{Hcf}
\end{equation}
with $\hn_{i\alpha}=\sum_\sigma\hn_{i\alpha\sigma}$ and $\hn_{i\alpha\sigma}=d^\dagger_{i\alpha\sigma}d_{i\alpha\sigma}$. In our case the electron energies
are given by $\epsilon_c=0$ for the $xy$ and $\epsilon_a=\epsilon_b=\Delta$ for the $xz$ and $yz$ orbitals. The electron-electron interactions
are described by the on-site terms,\cite{Oles83}

\begin{eqnarray}
\Ham_\textrm{int} &=& U\sum_{i\alpha}\hn_{i\alpha\uparrow}\hn_{i\alpha\downarrow}+\frac 12\left(U-\frac 52 J_H\right)\sum_{i\alpha\beta}^{\alpha\neq\beta}\hn_{i\alpha} \hn_{i\beta}\nonumber\\
&+&J_H \sum_{i\alpha\beta}^{\alpha\neq\beta}d^\dagger_{i\alpha\uparrow}d^\dagger_{i\alpha\downarrow}d_{i\beta\downarrow}d_{i\beta\uparrow} -J_H \sum_{i\alpha\beta}^{\alpha\neq\beta}\hat{\bS}_{i\alpha}\hat{\bS}_{i\beta},
\label{Hint}
\end{eqnarray}
with the Coulomb element $U$ and a Hund's exchange element $J_H$.

\subsection{Superexchange model}

In the limit of strong Coulomb repulsion, $t\ll U$, charge fluctuations $d_i^6d_j^6\rightleftharpoons d_i^7d_j^5$ are suppressed and on each site the two $t_{2g}$ electrons have to form a state belonging to the ground-state manifold of $\Ham_\textrm{int}+\Ham_\textrm{cf}$ in the two-electron sector. For sufficiently small crystal-field splitting, $\Delta^2<8 J_H^2$, these states are given by two $S=1$ triplets in which on each site either  the $xz$ or $yz$ is unoccupied. This orbital degree of freedom can be viewed as a $T=\frac 12$ pseudospin. From Eqs. (\ref{Hcf}), (\ref{Hint}) we easily obtain $E_0=U-3J_H+\Delta$ as the ground-state energy of the $t_{2g}^2$ sector.

A general spin-orbital superexchange model can be derived by second order perturbation theory controlled by the kinetic energy contribution $\Ham_t$, where we have to consider all virtual processes $t_{2g}^2t_{2g}^2\to t_{2g}^1t_{2g}^3\to t_{2g}^2t_{2g}^2$ acting on the $S=1$, $T=1/2$ ground-state manifold. The most general superexchange Hamiltonian in the sense of Kugel and Khomskii for a given bond $(i,j)$ takes the form 

\begin{eqnarray}
\Ham_{KK}^{(i,j)} & = &  -\sum_{\tau_i,\tau_j}\sum_{s_i,s_j} J_{\tau_i,\tau_j,s_i,s_j}^{(i,j)}A^{(i,j)}_{\tau_i,\tau_j}(\hat{\bT}_i,\hat{\bT}_j)\nonumber\\
& & \times B_{s_i,s_j}(\hat{\bS}_i,\hat{\bS}_j),
\label{HamKK}
\end{eqnarray}   
where $\hat{\bS}$ and $\hat{\bT}$ denote $S=1$ spin and $T=\frac 12$ pseudospin operators. The functional form of $B$ only depends on total spins $s_i$ and $s_j$ on the two sites in the intermediate $t_{2g}^1 t_{2g}^3$ states. Whereas the single occupied site has necessarily $s=1/2$ the other site can be in a high-spin ($s=3/2$) or low-spin ($s=1/2$) state. Likewise, the functions $A^{(i,j)}$ are determined by the pseudospins $\tau_i$, $\tau_j$ of the involved intermediate states.

To derive the effective spin-orbital superexchange model we first have to find the multiplet structure of the virtual intermediate $t_{2g}^3$ configurations. It is straightforward to diagonalize $\Ham_\textrm{cf}+\Ham_\textrm{int}$ (\ref{Hcf},\ref{Hint}) in the three-particle sector. The lowest energy we find for the $^4A_2$ quartet of $s=3/2$ high-spin intermediate states $|^4A_2,\frac 32,s^z\rangle$, with
$|s^z\rangle =|\frac 32\rangle  =  d^\dagger_{a\uparrow} d^\dagger_{b\uparrow} d^\dagger_{c\uparrow}|0\rangle$, 
$|\frac 12\rangle  =  \frac{1}{\sqrt{3}}(d^\dagger_{a\uparrow} d^\dagger_{b\uparrow} d^\dagger_{c\downarrow}+d^\dagger_{a\uparrow} d^\dagger_{b\downarrow} d^\dagger_{c\uparrow} +d^\dagger_{a\downarrow} d^\dagger_{b\uparrow} d^\dagger_{c\uparrow})|0\rangle$, 
$|-\frac 12\rangle  =  \frac{1}{\sqrt{3}}(d^\dagger_{a\downarrow} d^\dagger_{b\downarrow} d^\dagger_{c\uparrow}+d^\dagger_{a\downarrow} d^\dagger_{b\uparrow} d^\dagger_{c\downarrow} +d^\dagger_{a\uparrow} d^\dagger_{b\downarrow} d^\dagger_{c\downarrow})|0\rangle$ and 
$|-\frac 32\rangle  =  d^\dagger_{a\downarrow} d^\dagger_{b\downarrow} d^\dagger_{c\downarrow}|0\rangle$.
Their energy is $\epsilon(^4A_2)=E(^4A_2)-2E_0=U-3J_H$, where $E_0=U-3J_H+\Delta$ is the groundstate energy in the $t_{2g}^2$ sector. In order for the approach to be valid we have to assume that the system has a charge-transfer gap, $U-3J_H>0$ and that the hopping matrix elements are sufficiently small compared to the charge-transfer gap. All the other multiplets consist of intermediate $s=1/2$ doublets. The $^2E$ multiplet with excitation energy $\epsilon(^2E)=U$ consist of the two spin-$\frac 12$ doublets
\begin{eqnarray}
|^2E,\frac 12,\sigma\rangle_1 & = & \frac{1}{\sqrt{6}}(2d^\dagger_{a\sigma}d^\dagger_{b\sigma} d^\dagger_{c,-\sigma}
-d^\dagger_{a\sigma}d^\dagger_{b,-\sigma} d^\dagger_{c,\sigma}\nonumber\\
& & -d^\dagger_{a,-\sigma}d^\dagger_{b\sigma} d^\dagger_{c\sigma})|0\rangle\\
|^2E,\frac 12,\sigma\rangle_2 & = & \frac{1}{\sqrt{2}}(d^\dagger_{a,-\sigma}d^\dagger_{b\sigma} d^\dagger_{c\sigma}
-d^\dagger_{a\sigma}d^\dagger_{b,-\sigma} d^\dagger_{c\sigma})|0\rangle.\quad
\end{eqnarray}

Finally, we have multiplets $^2T_1^{(\Delta)}$, $^2T_2^{(\Delta)}$ which consist of spin-$\frac 12$ doublets and invoke doubly occupied orbitals,
\begin{equation}
|^2T_{1/2},\frac 12,\sigma\rangle  =  \frac{1}{\sqrt{2}}d^\dagger_{c\sigma}(d^\dagger_{a\uparrow} d^\dagger_{a\downarrow}\mp d^\dagger_{b\uparrow} d^\dagger_{b\downarrow})|0\rangle
\end{equation}
with excitation energies $\epsilon(^2T_1)=U$ and $\epsilon(^2T_2)=U+2J_H$ and
\begin{eqnarray}
|^2T_{1/2}^\Delta,\frac 12,\sigma\rangle_1 & = & d^\dagger_{a\sigma}(\sqrt{1-v_\mp^2}d^\dagger_{c\uparrow} d^\dagger_{c\downarrow}\mp v_\mp d^\dagger_{a\uparrow} d^\dagger_{a\downarrow})|0\rangle\\
|^2T_{1/2}^\Delta,\frac 12,\sigma\rangle_2 & = & d^\dagger_{b\sigma}(\sqrt{1-v_\mp^2}d^\dagger_{c\uparrow} d^\dagger_{c\downarrow}\mp v_\mp d^\dagger_{b\uparrow} d^\dagger_{b\downarrow})|0\rangle\qquad
\end{eqnarray}
with 
$
v_\mp   =  J_H/\sqrt{J_H^2+(\Delta\pm\sqrt{\Delta^2+J_H^2})^2}
$
and excitation energies
$
\epsilon(^2T_{1/2}^\Delta)=U+J_H\mp\sqrt{\Delta^2+J_H^2}.
$

The resulting charge-excitation spectrum is shown schematically in Fig.~\ref{level}c. Although the single occupied $t_{2g}^1$ site of a virtual $t_{2g}^1t_{2g}^3$ intermediate state gives no contribution to the on-site electron-electron interaction it can lead to an additional crystal-field energy $\Delta$ if the electron is in the $a$ or $b$ orbital.

Let us first focus on the purely magnetic parts $B_{s_i,s_j}(\hat{\bS}_i,\hat{\bS}_j)$ of the superexchange Hamiltonian, which can be determined entirely by group theoretical methods. To be precise, we consider a two-ion system in the state $|S_A,M_A\rangle\otimes  |S_B,M_B\rangle$ which can be classified by the total spin $S_t$ and the z-component $M_t$. Applying a hopping operator of the form
$
\Ham_t = -t \sum_\sigma(c^\dagger_{A\sigma}c_{B\sigma}+\textrm{h.c}),
$
which preserves the quantum numbers $S_t$ and $M_t$ because of the spin-rotation invariance we obtain an intermediate state $|s_A,m_A\rangle\otimes  |s_B,m_B\rangle$
with $s_A=S_A\pm 1/2$ and $s_B=S_B\pm 1/2$. The effective superexchange involving intermediate spins $s_A$, $s_B$ is given by the second order process 
\begin{equation}
E(S_t,s_A,s_B) = - \sum_{m_a,m_B}\frac{| \langle s_A m_A ,s_B m_B| \Ham_t |S_t M_t \rangle  |^2}{\Delta E}.\nonumber
\end{equation}
Using Clebsch-Gordan coefficients $C_{m_1m_2m}^{j_1j_2j}=\langle j_1j_2m_1m_2|jm\rangle$, we can express the total spin states as
\begin{equation}
|S_t M_t \rangle = \sum_{M_A,M_B}C_{m_Am_BM_t}^{S_AS_BS_t}|S_A,M_A\rangle\otimes  |S_B,M_B\rangle. \nonumber
\end{equation}
Since the operators $c_\sigma^\dagger$ and $c_\sigma$ are irreducible tensor operators of rank 1/2 we can use the Wigner-Eckart theorem to obtain
\begin{eqnarray}
\langle s_Am_A|c^\dagger_{A\sigma}|S_AM_A\rangle & = & \| c_A^\dagger\| C_{M_A\sigma m_A}^{S_A\frac 12s_A}\nonumber \\
\langle s_Bm_B|c_{B\sigma}|S_BM_B\rangle & = & \| c_B\| C_{M_B(-\sigma) m_B}^{S_B\frac 12s_B}(-1)^{\frac 12-\sigma}, \nonumber
\end{eqnarray} 
where we have used $\|\cdot\|$ as a short-hand notation for the reduced matrix elements. Using these expressions we can rewrite the exchange energy
as
$
E(S_t,s_A,s_B) = \frac{t^2}{\Delta E}(\| c_A^\dagger\|\cdot\| c_B\|)^2 B(S_t,s_A,s_B), 
$
where we can express the function $B$ in terms of a Wigner $6j$-symbol as
\begin{equation}
B(S_t,s_A,s_B)=-(2s_A+1)(2s_B+1)\begin{Bmatrix} S_A & s_A & \frac 12\\ s_B & S_B & S_t \end{Bmatrix}^2, \nonumber
\end{equation}
which by using the relation $S_t(S_t+1)=S_A(S_A+1)+S_B(S_B+1)+2\hat{\bS}_A\hat{\bS}_B$ can be simplified further to 
\begin{eqnarray}
B_{s_A,s_B} & = &  -\frac{2}{(2 S_A+1)(2 S_B+1)} \times\left\{(s_A+\frac 12)(s_B+\frac 12)\right.\nonumber\\
& & \left.\phantom{\frac 12}-\textrm{sign}[(s_A-S_A)(s_B-S_B)]\hat{\bS}_A\hat{\bS}_B\right\}.\qquad \nonumber
\end{eqnarray}
This expression we can evaluate for $S_A=S_B=S=1$ for the $s=3/2$ high- and $s=1/2$ low-spin intermediate states to obtain
the (normalized) spin-projection operators
\begin{eqnarray}
B_{\frac 32,\frac 12}(\hat{\bS}_i,\hat{\bS}_j)=-\frac 13(\hat{\bS}_i\hat{\bS}_j+2)\\
B_{\frac 12,\frac 12}(\hat{\bS}_i,\hat{\bS}_j)=\frac 13(\hat{\bS}_i\hat{\bS}_j-1)
\end{eqnarray}
in agreement with Refs. \onlinecite{Oles+05,Oles+07}. Hence, the Kugel-Komskii superexchange Hamiltonian for a given bond $(i,j)$ can be written as,
\begin{eqnarray}
\Ham_{KK}^{(i,j)} & = & -\frac{1}{3}(\hat{\bS}_i\hat{\bS}_j+2)\mathcal{Q}^{(1)}(\hat{\bT}_i,\hat{\bT}_j)\nonumber\\
& & +\frac{1}{3}(\hat{\bS}_i\hat{\bS}_j-1)\mathcal{Q}^{(2)}(\hat{\bT}_i,\hat{\bT}_j) ,
\end{eqnarray}
where $\mathcal{Q}^{(n)}$ are functions of orbital pseudospin operators. Their functional
form can be obtained by tracking the orbital occupancies in the initial and final states
during a virtual hopping process. In terms of spinless Fermi operators $a_i^+$, $b_i^+$ increasing the occupancy of the $a$ or $b$ 
orbital on site $i$ the pseudospin-$1/2$ operators 
acting on the ground-state manifold can be expressed as $\hat{T}_i^z=(\hat{n}_{ia}-\hat{n}_{ib})/2$, $\hat{T}_i^+=b_i^+a_i$,
and $\hat{T}_i^-=a_i^+b_i$, where $\hat{n}_{ia}=a_i^+a_i$ and  $\hat{n}_{ib}=b_i^+b_i$ with the constraint $\hat{n}_{ia}+\hat{n}_{ib}=1$.
Whereas it is straightforward to see that the general functional form is given by

\begin{eqnarray}
& &\mathcal{Q}^{(n)}(\hat{\bT}_i,\hat{\bT}_j) = f_{zz}^{(n)}\hT_i^z\hT_j^z 
+\frac 12 f_{+-}^{(n)}(\hT_i^+\hT_j^-+\hT_i^-\hT_j^+) \nonumber \\
& &+\frac 12 f_{++}^{(n)}(\hT_i^+\hT_j^++\hT_i^-\hT_j^-) 
+f_{zx}^{(n)}(\hT_i^z\hT_j^x+\hT_i^x\hT_j^z) \nonumber \\
& &+f_{z}^{(n)}(\hT_i^z+\hT_j^z)+f_{x}^{(n)}(\hT_i^x+\hT_j^x)  +f_{0}^{(n)},
\end{eqnarray}
it is quite tedious to determine the coefficients by acting with the hopping operator $\Ham_t$ (\ref{Ht}) on all 
states in the ground state sector and calculating the overlap of the resulting states projected on the different intermediate states 
listed above. The resulting explicit expressions are given in Appendix A.

\subsection{Hopping and Resulting Hamiltonian}

\begin{figure}
\includegraphics[width=0.9\linewidth, clip = true]{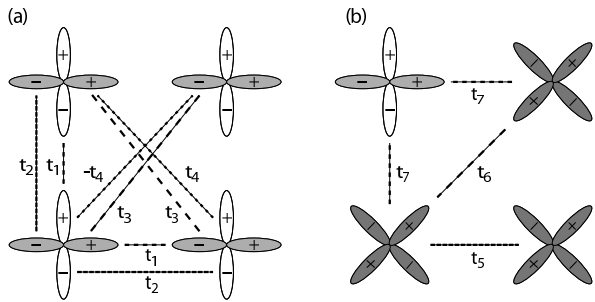}
\caption{Illustration of the effective hopping parameters $t_{\alpha\beta}$,  (a) between the
$d_{xz}$ and $d_{yz}$ orbitals, and (b) those involving the $d_{xy}$ orbitals.
The projections of the $d_{xz}$ and $d_{yz}$ orbitals on the Fe-plane are depicted in white and
light grey, respectively, and the $d_{xy}$ orbitals are shown in dark grey}
\label{Fe_hopping}
\end{figure}

In the previous section we have derived the general KK superexchange Hamiltonian only assuming the effective hopping matrices to be
symmetric, $t_{\alpha\beta}=t_{\beta\alpha}$. In order to write down the spin-orbital model specific to the pnictide planes we have to use the corresponding hopping parameters. We use the Slater-Koster integrals\cite{slaterkoster} along with the geometry of the Fe-As planes to determine all the hopping parameters involving the three $t_{2g}$ orbitals on the nearest neighbor and next nearest neighbor Fe sites.
This considerably reduces the number of independent hopping parameters that enter the Hamiltonian. The direct $d-d$ hoppings are considered to be much smaller therefore we use hoppings via the As-$p$ orbitals only which are given in Appendix B and depend on the direction cosines $l,m,n$ of the As-Fe bond, as well as on the ratio $\gamma=(pd\pi)/(pd\sigma)$\cite{Scalapino,Hirschfeld,Dagotto}. These resulting effective hopping matrix elements between the $t_{2g}$ Fe orbitals are shown schematically in Fig.~\ref{Fe_hopping} and can be parametrized by the lattice parameter $\lambda=|n/l|$ and $\gamma$.

\begin{figure}
 \includegraphics[width=0.85\linewidth, clip = true]{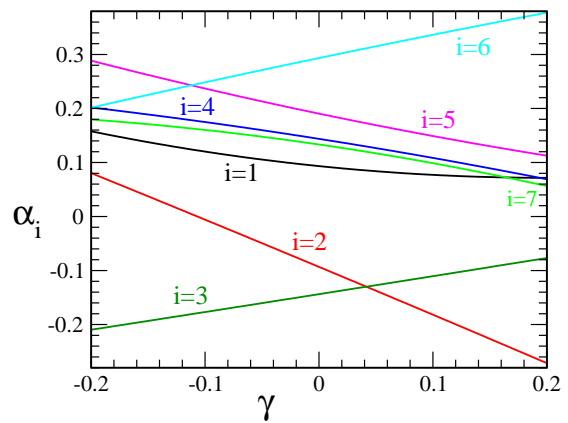}
 \caption{ (Color online) Various hopping parameters $\alpha_i:=t_i/t$ as illustrated in Fig.~\ref{Fe_hopping} as a function of the ratio $\gamma = (pd\pi)/(pd\sigma)$
for the lattice parameter $\lambda = 0.7$.}
\label{t_vs_gamma}
\end{figure}

In Fig.~\ref{t_vs_gamma} the dependence of the hopping matrix elements on the ratio $\gamma = (pd\pi)/(pd\sigma)$ is shown 
for a lattice parameter $\lambda=0.7$ which is slightly below the value resulting from the Fe-Fe spacing and the distance of the As ions
to the Fe planes. Over a realistic range $-0.2\le\gamma\le 0.2$ we find a very strong dependence of the hopping amplitudes 
on $\gamma$ and therefore expect the stability of possible phases to depend crucially on $\gamma$. This parameter cannot be obtained by 
geometrical considerations but depends for instance on how strongly the orbitals delocalize. 

Having specified the effective hopping parameters $\alpha_i:=t_i/t$ between the Fe orbitals for nearest and next-nearest neighbors (see Fig.~\ref{Fe_hopping})
which are parametrized entirely by the ratio $\gamma = (pd\pi)/(pd\sigma)$ and the lattice parameter $\lambda=|n/l|$
we can now write down the effective KK model for the Fe planes. For convenience, we rewrite the Hamiltonian in the form

\begin{equation}
\Ham_{KK} = J\sum_{(i,j)}\left(\frac 12(\hat{\bS}_i\hat{\bS}_j+1)\hat{\Omega}_{(i,j)}+ \hat{\Gamma}_{(i,j)}\right)
\end{equation}
introducing an overall energy scale $J=4t^2/U$. The orbital bond operators are defined as $\hat{\Omega}=\frac{U}{6t^2}(\mathcal{Q}^{(2)}-\mathcal{Q}^{(1)})$ and $\hat{\Gamma}=-\frac{U}{12t^2}(\mathcal{Q}^{(1)}+2\mathcal{Q}^{(2)})$ and depend on the effective couplings $\alpha_i$, the relative strength of the Hunds coupling $\eta=J_H/U$ and the crystal-field splitting $\delta=\Delta/U$. For the nearest neighbor bonds along $\hat{x}$ and $\hat{y}$ along the $\hat{x}\pm\hat{y}$ diagonals the operators are given in Appendix C.

\section{Classical phase diagrams}

In this section we discuss the phase diagrams of the spin-orbital Hamiltonian in the classical limit. We have four parameters that enter the model: $\lambda$ and $\gamma$ determine the relative strength of various hopping parameters and $\eta$ and $\delta$ enter via the energy denominators. Zero temperature phase transitions are discussed in subsection A, subsection B is devoted to the understanding of finite temperature transitions and subsection C analyses the phases in terms of the corresponding spin-only and orbital-only models.

The results that we discuss below demonstrate that the Hamiltonian is highly frustrated in the spin and orbital variables. While the spin frustration is largely due to the competing interactions between nearest- and next-nearest neighbors, the frustration in orbital sector is more intrinsic and exists within a single bond in the Hamiltonian. The spin ($\pi,0$) state is found to be stable over a wide range of parameter space due to the strong nnn AF coupling. However, depending on the parameters, there are three possible orderings of the orbitals that accompany the spin 'stripe' order. Two out of these three orbital ordering patterns break the in-plane symmetry of the lattice and hence are likely candidates for explaining the orthorhombic transition observed in the parent compounds.

\subsection{Zero temperature}

Since the effective KK Hamiltonian derived in section \ref{sec.KK} contains a large number of competing terms it is almost impossible to anticipate what kind of spin-orbital orderings are realized for different parameter values, in particular since the signs and relative strengths of the effective hoppings $\alpha_i$ between nearest and next-nearest neighbor Fe orbitals crucially depend on the ratio $\gamma = (pd\pi)/(pd\sigma)$ as pictured in Fig.~\ref{t_vs_gamma}. While the parameters $\alpha_1$, $\alpha_4$ and $\alpha_7$ do not show large relative changes over the range of $\gamma$ shown in the figure, there are very clear crossings between $\alpha_2$ and $\alpha_3$ and $\alpha_5$ and $\alpha_6$.

\begin{figure}
 \includegraphics[width=.9\linewidth,clip = true ]{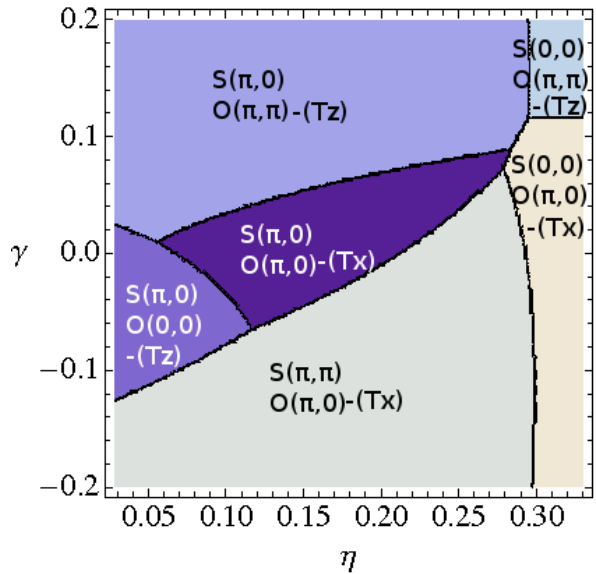}
 \caption{ (Color online) $\eta$-$\gamma$ phase diagram for $\lambda = 0.7$ and $\delta=0.01$. $\eta = J_H/U$, $\gamma = (pd\pi)/(pd\sigma)$. The phases are denoted by their ordering wavevectors in the
spin and orbital variables. $Tz$ or $Tx$ refers to the component of the orbital pseudospin that is
saturated in the ordered state.}
\label{pd_eta_gamma}
\end{figure}

Recall that $\alpha_5$ and $\alpha_6$ are the hoppings between nearest and next-nearest neighbors involving
orbital $|c\rangle:=|xy\rangle$. If we infer the spin order arising purely from the non-degenerate $|c\rangle$ orbital, it suggests that
the spin state should be $(\pi,0)$-ordered for $\alpha_5^2 < 2\alpha_6^2 $ and $(\pi,\pi)$-ordered otherwise. Therefore, this
would imply that as $\gamma\to -0.2$ the magnetic superexchange resulting from the $|c\rangle$ orbitals only favors
$(\pi,\pi)$ antiferromagnetism, whereas the $(\pi,0)$ stripe AF becomes favorable for $\gamma\to 0.2$. 

A similar spin-only analysis for the degenerate orbitals $|a\rangle$,$|b\rangle$ is not possible and one has to treat the
full spin-orbital Hamiltonian in order to find the groundstates. Nevertheless, the complicated variations in
the hopping parameters already suggest that we can expect a very rich and complex phase diagram for the
groundstate of the spin-orbital Hamiltonian. In particular in the region of intermediate $\gamma$ where the magnetic superexchange
model resulting form the $|c\rangle$ orbitals only becomes highly frustrated we expect the magnetic ordering to depend 
crucially on the orbital degrees of freedom.

We first look at the classical groundstates of this model. We make use of classical Monte-Carlo
method in order to anneal the spin and orbital variables simultaneously, starting with a completely random
high temperature configuration. Using this method we identify the various groundstates that exist for a combination of model 
parameters. In order to obtain a groundstate phase diagram, we minimize the total
energy for a set of variational states which also include all the Monte-Carlo groundstates obtained
for different choice of parameters.

\begin{figure}
 \includegraphics[width=\linewidth,clip = true ]{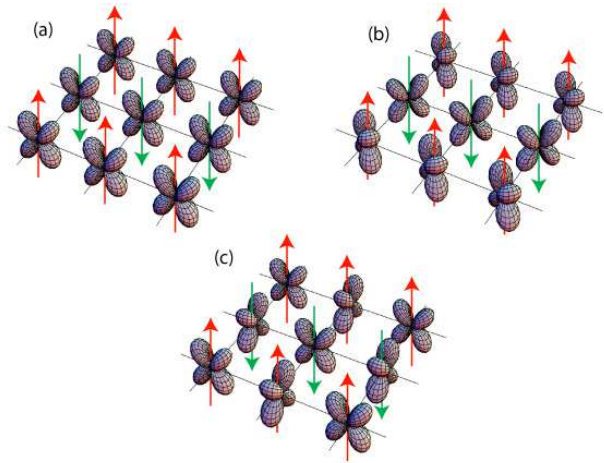}
 \caption{ (Color online) Schematic pictures of the three ground-state orbital ordering patterns that
accompany the spin-stripe phase.
(a) Orbital-ferro, (b) orbital-stripe, and (c) orbital-antiferro }
\label{orbitals}
\end{figure}

Fig.~\ref{pd_eta_gamma} shows the resulting $T=0$ phase diagram for varying $\eta = J_H/U$ and $\gamma = (pd\pi)/(pd\sigma)$. The lattice parameter $\lambda$
is fixed to $0.7$, which is close to the experimental value for the oxypnictides. The crystal-field splitting between the $|c\rangle$ and
the $|a\rangle$,$|b\rangle$ orbitals is considered to be very small, $\delta=\Delta/U=0.01$. As expected, a large number of phases are present in 
the phase diagram.

With increasing $\gamma$ we indeed find a transition from a $(\pi,\pi)$ to a $(\pi,0)$ antiferromagnet as suggested from the analysis of the
frustrated magnetic superexchange model involving only the $|c\rangle$ orbitals. This is not surprising since the corresponding couplings $\alpha_5^2$ and/or
$\alpha_6^2$ are sufficiently strong and as $\gamma\to 0.2$ the biggest hopping element is in fact given by $\alpha_6$ between next-nearest neighbor $|c\rangle$ 
orbitals (see Fig.~\ref{t_vs_gamma}). Whereas the $(\pi,0)$ stripe magnet for large $\gamma$ is accompanied by an antiferro-orbital ordering of the $T_z$
components corresponding to a checkerboard arrangement of the $|a\rangle$ and $|b\rangle$ orbitals (see Fig.~\ref{orbitals}c) for intermediate, small 
$\gamma$ we find two $(\pi,0)$ magnetic phases possessing orbital orderings which are likely to break the inplane symmetry of the lattice structure.

For small $\eta$ we find a ferro-orbital arrangement of the $T_z$ components corresponding to the formation of chains along the ferromagnetically
coupled spin directions (see Fig.~\ref{orbitals}a). The existence of this orbital order crucially depends
on the pre-existence of a spin stripe state, which generates magnetic-field-like terms for the orbital
pseudospins. This will be discussed in detail when we try to understand the thermal phase transitions.
For larger $\eta$ the orbital order changes to an orbital-($\pi,0$) 'tweed' pattern with a condensation of the $T_x$
components. This corresponds to the formation of orbital zig-zag chains along the antiferromagnetically coupled spin direction as pictured in Fig.~\ref{orbitals}b.
Interestingly, the stripes in the magnetic and orbital sectors have the same orientation, contrary to the conventional Goodenough-Kanamori rules. However, since we are dealing with a highly frustrated spin-orbital model involving nearest and next-nearest neighbor bonds these naive rules are not expected to hold. The 'tweed' orbital order is expected to lead to a displacement pattern of the As ions, which can in principle be observed in X-ray differaction experiments. The 'tweed' orbital pattern should show up as a higher order structural Bragg peak at ($\pi,0$). The orbital order might also be directly visible resonant X-ray diffraction at the iron K-edge, a technique that was pioneered in the manganites~\cite{Murakami98,Elfimov99,Benedetti01},  and is nowadays available for all transition metal K-edges, in particular the iron one~\cite{Joly08}. Polarization analysis and azimuthal angle dependence can distinguish between charge, spin and orbital contributions to the resonant signal~\cite{Murakami98} which gives the possibility in the iron pnictides to single out the 'tweed' orbital pattern.

\begin{figure}
 \includegraphics[width=0.9\linewidth, clip = true]{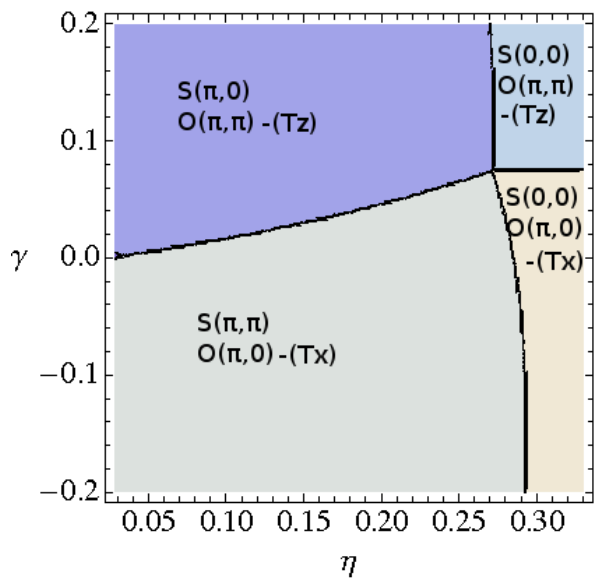}
 \caption{(Color online) $\eta$-$\gamma$ phase diagram for $\lambda = 0.8$ and $\delta=0.01$. Note that
the orbital ordered states that break the orthorhombic symmetry do not exist for this choice of $\lambda$.}
\label{pd2_eta_gamma}
\end{figure}

The orbital stripe order persists to the regime of larger negative $\gamma$ where the magnetic order changes to the $(\pi,\pi)$ antiferromagnet. This shows that the orbital 'tweed' state does not have spin-$(\pi,0)$ order as a pre-requisite and therefore this orbital order can, in principle, exist at temperatures higher than the spin transition temperatures. In the regime of large Hund's coupling, $\eta \ge 0.3$ the system becomes ferromagnetic. This tendency is easy to understand since in the limit $\eta\to1/3$ the charge-transfer gap closes and the KK model is dominated by  processes involving the low-lying $^4A_2$ high-spin multiplet favoring a ferromagnetic superexchange. 

Let us further explore how the groundstate phase diagram changes as we vary the lattice parameter $\lambda$ and the crystal-field splitting $\delta$. 
Fig.~\ref{pd2_eta_gamma} shows the same phase diagram as in Fig.~\ref{pd_eta_gamma} but for a slightly larger separation of the As ions to the Fe-planes,
$\lambda = 0.8$. The two interesting phases with magnetic $(\pi,0)$ and orbital-stripe and orbital ferro orderings do not appear in this phase diagram indicating that the
stability of these phases crucially depends on the relative strength of nearest and next-nereast hoppings which can be tuned by $\lambda$. Presence
of a tetracritical point is an interesting feature in this phase diagram.

\begin{figure}
 \includegraphics[width=0.9\linewidth, clip = true]{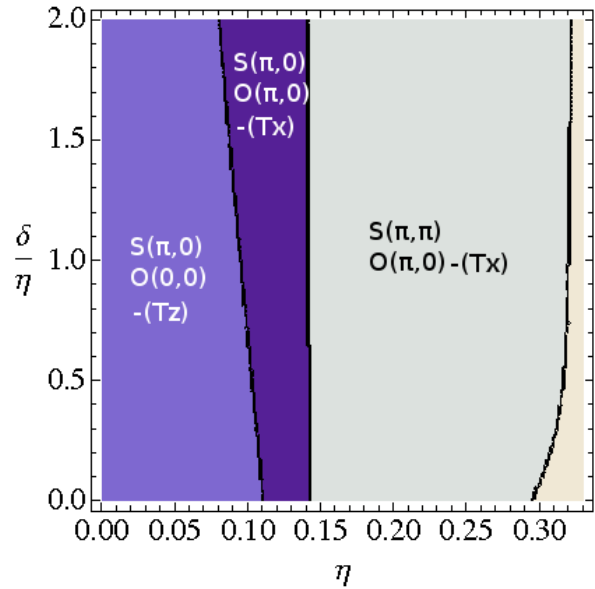}
 \caption{(Color online) $\eta$-$\delta/\eta$ phase diagram for $\lambda = 0.7$ and $\gamma =
(pd\pi)/(pd\sigma)= -0.05$. This phase diagram illustrates the point that $\delta$ is not a crucial
parameter in the Hamiltonian.}
\label{pd_eta_delta}
\end{figure}

Finally, we analyze the dependence on the crystal field splitting $\delta=\Delta/U$ which so far we assumed to be tiny. We do not find any qualitative
change of the groundstate phase diagram with increasing $\delta$. In particular, there are no new phases that appear and therefore
the crystal-field splitting does not seem to be a crucial parameter. For example, the phase diagram in the 
$\eta$-$\delta/\eta$-plane for $\lambda = 0.7$ and $\gamma = -0.05$ shown in Fig.~\ref{pd_eta_delta} indicates that a change in $\delta$ only
leads to a small shift of the phase boundaries.

\subsection{Finite temperature}

To obtain the transition temperatures for the various phase transitions, we track different order parameters as a function of temperature during Monte-Carlo annealing where we measure the temperature in units of the energy scale $J$. For example, the spin structure factor is defined as

\begin{equation}
S({\bf q}) = \frac{1}{N^2} \sum_{i,j} \langle {\bf S}_i \cdot {\bf S}_j \rangle_{av}~e^{i{\bf q} \cdot ({\bf r}_i - {\bf r}_j)},
\end{equation}
where $\langle ... \rangle_{av}$  denotes thermal averaging and $N$ is the total number of lattice sites. The orbital structure factor $O({\bf q})$ is defined analogously by
replacing the spin variables by the orbital variables in the above expression. Depending on the groundstate, different components of these structure factors
show a characteristic rise upon reducing temperature.

We fix $\delta = 0.01$, $\lambda=0.7$ and $\gamma = -0.05$ and track the temperature dependence of the
system for varying $\eta$. For $T=0$ this choice of parameters corresponds to a cut of the phase diagram shown in Fig.~\ref{pd_eta_gamma} through four 
different phases including the two $(\pi,0)$ stripe AFs with orbital orderings breaking the in-plane lattice symmetry.

\begin{figure}
 \includegraphics[width=.9\linewidth, clip = true]{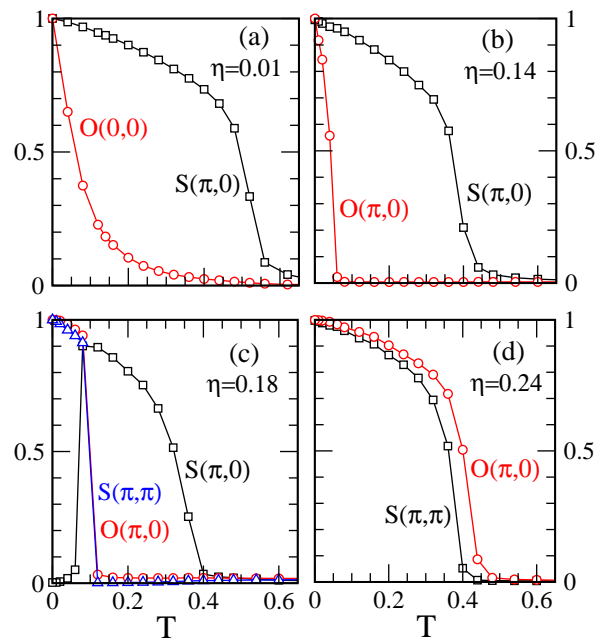}
 \caption{(Color online) Relevant structure factors as a function of temperature for different values of $\eta$. The lattice parameter and the relative strength of $\sigma$ and $\pi$ hopping are fixed as $\lambda = 0.7$ and $\gamma = -0.05$, respectively.}
\label{T_eta}
\end{figure}

In Fig.~\ref{T_eta} the temperature dependence of the corresponding structure factors is shown for representative values of the Hund's rule coupling $\eta$.
For small values of $\eta$ the groundstate corresponds to the orbital-ferro and spin-stripe state as shown in the phase diagram in Fig.~\ref{pd_eta_gamma}.
Fig.~\ref{T_eta}a shows the temperature dependence of $S(\pi,0)$ and $O(0,0)$ which are the order parameters
for the spin-stripe and orbital-ferro state, respectively. While the $S(\pi,0)$ leads to a characteristic curve with the steepest rise at $T \sim 0.5$, the
rise in $O(0,0)$ is qualitatively different. In fact there is no transition at any finite $T$ in the orbital
sector. We can still mark a temperature below which a significant
orbital-ferro ordering is present. The origin of this behavior lies in the presence of a Zeeman-like term
for the orbital pseudospin.

For $\eta = 0.15$, the phase diagram of Fig.~\ref{pd_eta_gamma} suggests a state with stripe ordering in both spin and orbital
variables. We show the temperature dependence of $S(\pi,0)$ and $O(\pi,0)$ in Fig.~\ref{T_eta}b. In this case both the spin and orbital variables show a spontaneous ordering, with the spins ordering at a much higher temperature. 
An interesting sequence of transitions
is observed upon reducing temperature for $\eta = 0.18$ (see Fig.~\ref{T_eta}c). This point lies close to the phase boundary between spin-stripe and spin-ferro state with the orbital-stripe ordering. The spin-stripe order parameter $S(\pi,0)$ shows a strong rise near $T=0.4$.
The orbital stripe order sets in at $T \sim 0.15$. The onset of this orbital order kills the
spin-stripe order. Instead, we find that the $(\pi,\pi)$ components of the spin structure factor
shows a strong rise along with the $(\pi,0)$ component of the orbital structure factor.
Finally for $\eta = 0.24$ the orbital stripe ordering is accompanied by the spin antiferro ordering,
with the orbital ordering setting in at slightly higher temperatures (see Fig.~\ref{T_eta}d).

\begin{figure}
 \includegraphics[width=0.85\linewidth, clip = true]{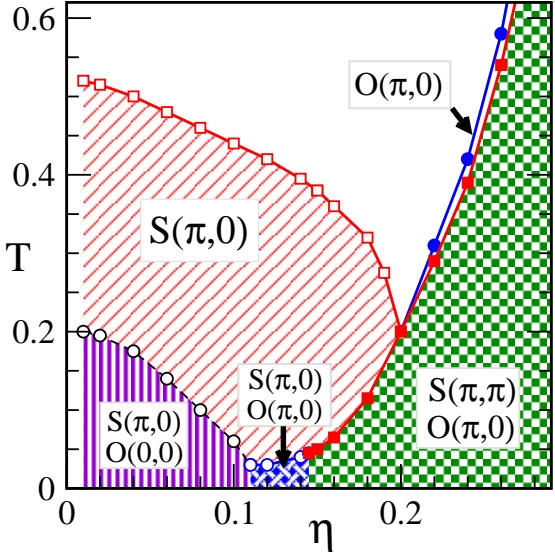}
 \caption{(Color online) $T$-$\eta$ phase diagram for $\delta=0.01$, $\lambda = 0.7$, and $\gamma = -0.05$}
\label{T_eta2}
\end{figure}

The results shown in Fig.~\ref{T_eta} are summarized in the $T-\eta$ phase diagram shown in Fig.~\ref{T_eta2}.
For small $\eta$, the groundstate is spin-stripe and orbital-ferro ordered. While the
spin order occurs at higher temperatures, there is no genuine transition to the
orbital-ferro state. The orbital-ferro state is driven by the presence of a magnetic-field-like
term for the orbital pseudospin in the Kugel-Khomskii Hamiltonian. The stability of the orbital-ferro
state crucially depends on the presence of the spin-stripe order. The dotted line joining the black
circles in the small-$\eta$ range is only to indicate the temperature below which the orbital-ferro order
is significant. This typical temperature scale reduces with increasing $\eta$, until the system finds a different groundstate for the orbital variables. Note that the temperature scales involved are very small
owing to the highly frustrated nature of the orbital model, nevertheless there is no zero-temperature
transition in this purely classical limit.

The spin-stripe state remains stable with the transition temperature reducing slightly.
The transition temperature for the orbital stripe state increases upon further increasing
$\eta$. For $0.15 < \eta < 0.2$, multiple thermal transitions are found for the magnetic state. The spin-stripe order which sets in nicely at $T \sim 0.35$ is spoiled by the onset of orbital-stripe state, which instead stabilizes the spin $(\pi,\pi)$ state. Beyond $\eta = 0.2$,
The orbital stripe state occurs together with the spin antiferro state, with the spin ordering temperatures
slightly lower than those for the orbital ordering. For $\eta > 0.3$, the spin state becomes ferromagnetic.

\subsection{Corresponding Orbital-only and Spin-only models}
\begin{figure}
 \includegraphics[width=0.9\linewidth, clip = true]{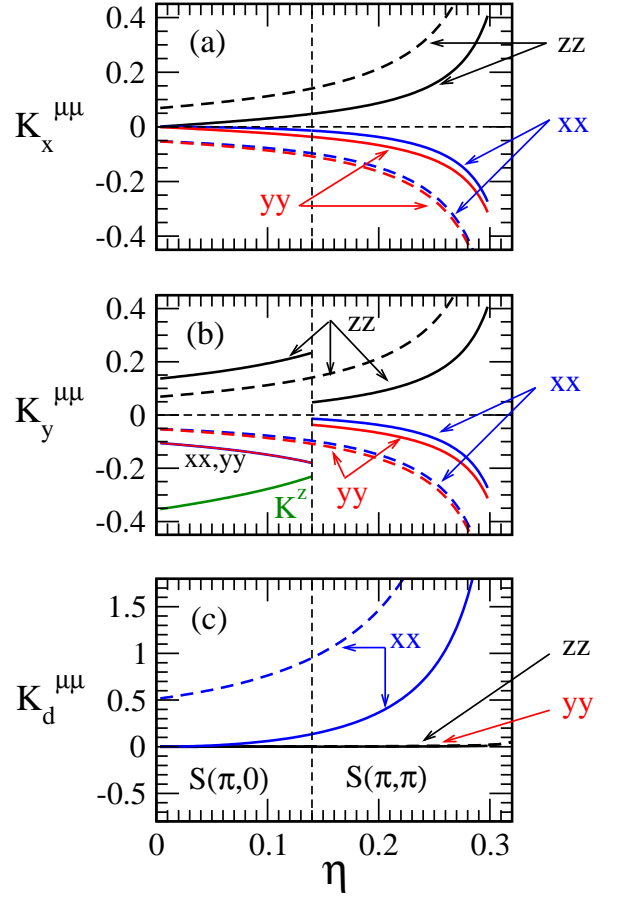}
 \caption{(Color online) The coupling constants as a function of $\eta=J_H/U$ for the orbital-only 
model with frozen spin correlations
for $\delta=0.01$, $\gamma=-0.05$, and $\lambda=0.7$.
The couplings along x, y and diagonal directions are plotted in panels (a), (b) and (c) respectively.
The single site term is plotted in (b) to indicate that this term arises due to a ferromagnetic bond along
y-direction.
The solid lines correspond to the ground state spin order and the dashed lines are for a paramagnetic spin state. The vertical dashed line indicates the location in $\eta$ of the phase transition from spin-stripe to
spin-antiferro state as seen in
Fig.~\ref{T_eta2}}
\label{KxKy}
\end{figure}

In an attempt to provide a clear understanding of the spin and orbital ordered phases, we derive the orbital (spin) model that emerges by freezing the spin (orbital) states. For fixed spin correlations, the orbital model can be written as

\begin{eqnarray}
\Ham_{O} & = & \sum_{\mu} K_x^{\mu \mu} \sum_{\langle i,j \rangle\parallel x} T_i^{\mu} T_j^{\mu}
+ \sum_{\mu} K_y^{\mu \mu} \sum_{\langle i,j \rangle\parallel y} T_i^{\mu} T_j^{\mu} \nonumber  \\
& + & \sum_{\mu} K_d^{\mu \mu} \sum_{\llangle i,j \rrangle} T_i^{\mu} T_j^{\mu}
+ K^z \sum_i T_i^z.
\end{eqnarray}
Here and below $\langle \cdot,\cdot\rangle$ and  $\llangle \cdot,\cdot\rrangle$ denote bonds between nearest and next-nearest neighbor pseudospins on the square lattice, respectively. $\mu$ denotes the component of the
orbital pseudospin. The effective exchange couplings for this orbital-only model are shown in
Fig.~\ref{KxKy} as a function of the Hund's coupling $\eta=J_H/U$ with the other parameters fixed as $\delta=0.01$, $\gamma=-0.05$, and $\lambda=0.7$, as before. The solid lines
are obtained by fixing the spin degrees of freedom by the classical ground-state configurations of the corresponding phases. For comparison, the effective couplings
for disordered spins are shown by dashed lines.

\begin{figure}
 \includegraphics[width=0.9\linewidth, clip = true]{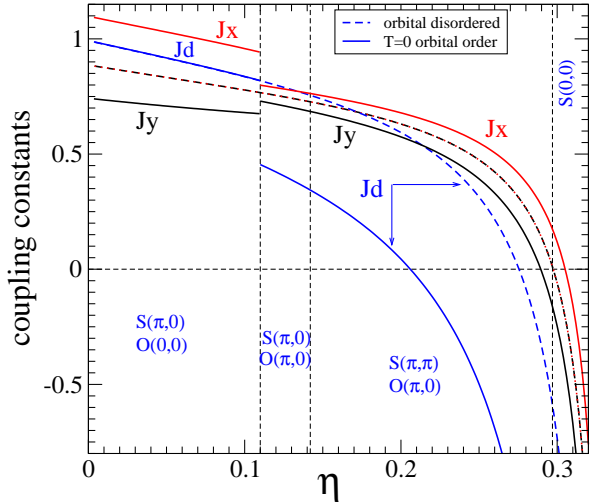}
 \caption{(Color online) Effective exchange couplings $J_x$, $J_y$ for nearest-neighbor and $J_d$ for next-nearest-neighbor spins as a function of $\eta=J_H/U$
 for $\delta=0.01$, $\gamma=-0.05$, and $\lambda=0.7$. The solid lines correspond to the couplings resulting for the corresponding orbital ground states whereas the
 dashed lines correspond to the orbitally disordered case. Note that for the orbitally disordered case $J_x = J_y$ for all values of $\eta$. The vertical dashed line indicates the location in $\eta$ of the various phase transition as seen in Fig.~\ref{T_eta2}.}
\label{Js}
\end{figure}

Similarly, we can freeze the orbital degrees of freedom to obtain an effective Heisenberg model for
spins,
\begin{equation}
\Ham_{S} = J_x\sum_{\langle i,j\rangle\parallel x}\bS_i\bS_j+J_y\sum_{\langle i,j\rangle\parallel y}\bS_i\bS_j+J_d\sum_{\llangle i,j\rrangle}\bS_i\bS_j.
\label{J1J2}
\end{equation}
The coupling constants $J_x$, $J_y$, and $J_d$ for spins are plotted in Fig.~\ref{Js}.

Let us try to understand the phase diagram of Fig.~\ref{T_eta2} in terms of these coupling constants.
We begin with the small-$\eta$ regime where the groundstate is spin-stripe and orbital-ferro.
Approaching from the high temperature limit, we should look at the spin (orbital) couplings with
disordered orbitals (spins). The strongest constants turn out to be $J_d$, which is slightly larger than
$J_x$ and $J_y$, all three being antiferromagnetic. This suggests that the system should undergo a transition to a spin-stripe state consistent
with the phase diagram. The coupling constants of the orbital model are much weaker in the small-$\eta$
regime. The largest constant is $K_d^{xx}$ suggesting an orbital-stripe order.
However, since the spin-
stripe state sets in at higher temperatures, in order to determine the orbital order one should look at the coupling constants corresponding to the spin-stripe state.
There are three main effects (compare the solid and dashed lines in the low $\eta$ regime in
Fig.~\ref{KxKy}), (i) x- and y-directions become inequivalent in the sense that the couplings along x are
suppressed while those along y are enhanced, (ii) the
diagonal couplings are reduced strongly, and (iii) a
single-site term is generated which acts as magnetic field for the orbital pseudospins. It is in fact this
single site term that controls the ordering of the orbitals at low temperatures. This also explains the
qualitatively different behavior of the orbital-ferro order parameter observed in Fig.~\ref{T_eta}a.
Within the spin-stripe order,
the single-site term becomes weaker with increasing $\eta$ whereas the diagonal term increases. This leads
to a transition in the orbital sector from an orbital-ferro to an orbital-stripe phase near $\eta=0.11$.
The region between $0.14$ and $0.2$ in eta is very interesting. Approaching from the high temperature
the spins order into the 'stripe' state but as soon as the orbitals order into stripe state at lower temperature
the diagonal couplings $J_d$ are strongly reduced and become smaller than $J_y/2$. This destabilizes the spin-stripe state and leads to
an spin antiferro ordering. For larger $\eta$ the orbital ordering occurs at higher temperature. There
is another transition slightly below  $\eta=0.3$ where spins order into a ferro state. This is simply understood as $J_x = -J_y$ from the coupling constants of the Heisenberg model.

\section{Magnetic excitation spectra}

We now set out to compute the magnetic excitation spectra, treating the orbital pseudospins as classical and static variables. Fixing the orbital degrees of freedom for a given set of parameters by the corresponding ground-state configuration we are left with an $S=1$ Heisenberg model written in (\ref{J1J2}). The exchange couplings are plotted in Fig.~\ref{Js}.
%
Assuming the presence of local moments, such $J_1$-$J_2$ models with a sufficiently large next-nearest neighbor exchange have been
motivated and used to rationalize the $(\pi,0)$ magnetism in the iron pnictides\cite{Si+08} and been used subsequently to calculate the magnetic excitation 
spectra\cite{Yao+08,Zhao+08}, where the incorporation of a relatively strong anisotropy between the nearest-neighbor couplings turned out to be necessary to understand the low energy spin-wave excitations\cite{Zhao+08}.

In the presence of orbital ordering such an anisotropy of the effective magnetic exchange couplings appears naturally. Both, the orbital ferro and the orbital stripe order lead to a seizable anisotropy in the nearest-neighbor couplings $J_x$, $J_y$, where the anisotropy is much stronger for the orbital ferro order (see Fig.~\ref{Js}). An even more drastic effect is the huge suppression of $J_d$ in the orbital-stripe regime.

\begin{figure}
 \includegraphics[width=0.7\linewidth, clip = true]{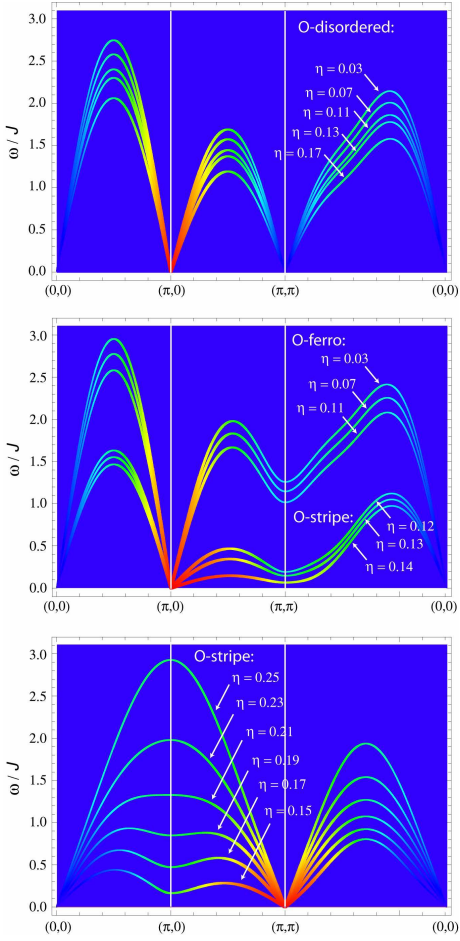}
 \caption{(Color online) Spin-wave excitation spectra for different values of $\eta$. Top: $(\pi,0)$ magnet for disordered orbitals. The spectral weights are coded by line 
 thickness and color, high intensity corresponds to red, low intensity to blue. Middle: $(\pi,0)$ magnet for orbital-ferro 
 ($\eta=0.03, 0.07, 0.11$) and orbital-stripe order ($\eta=0.12, 0.13, 0.14$). Bottom: $(\pi,\pi)$ magnet with orbital-stripe order ($\eta=0.15, 0.17, \ldots$).}
\label{swave}
\end{figure}

On a classical level, the magnetic transitions are easily understood in the spin-only model (\ref{J1J2})
as discussed before. The transition from the stripe-AF to the $(\pi,\pi)$ AF at $\eta\approx 0.14$
occurs exactly at the point where $J_y=2J_d$ whereas the transition from $(\pi,\pi)$ to ferromagnetic order at $\eta\approx 0.3$ corresponds to the point $J_x=-J_y$. 

We proceed to calculate the magnetic excitation spectra in the $\bQ=(\pi,0)$ and $(\pi,\pi)$ phases within a linear spin-wave approximation. The classical ground states are given given by $\bS_\br=S(0,0,\sigma_\br)$ with $\sigma_\br=\exp(i\bQ\br)=\pm 1$. After performing a simple spin rotation, $S^x=\tilde{S}^x_\br$, $S^y_\br=\sigma_\br\tilde{S}^y_\br$, and $S^z_\br=\sigma_\br\tilde{S}^z_\br$, we express the rotated spin operators by Holstein-Primakoff bosons, $\tilde{S}^+=\sqrt{2S-\hat{n}}b$,
$\tilde{S}^-=b^\dagger\sqrt{2S-\hat{n}}$, and $\tilde{S}^z=S-\hat{n}$ with $\hat{n}=b^\dagger b$ to obtain the spin-wave Hamiltonian
\begin{equation}
\Ham = S\int_\bq\left\{A_\bq(b_\bq^\dagger b_\bq  + b_{-\bq}b^\dagger_{-\bq})+B_\bq(b_\bq^\dagger b^\dagger_{-\bq}+b_{-\bq} b_{\bq})  \right\}, \nonumber
\label{Hsw}
\end{equation}
with 
\begin{eqnarray}
A_\bq &=& \left[ -J_x\cos Q_x+J_x\frac{1+\cos Q_x}{2}\cos q_x \right. \nonumber \\ 
& & -J_d\cos Q_x\cos Q_y\nonumber\\
& & \left. +\frac {J_d}{2}(1+\cos Q_x\cos Q_y)\cos q_x\cos q_y \right] + x \leftrightarrow y \nonumber \\
B_\bq & = & J_x\frac{1-\cos Q_x}{2}\cos q_x+J_y\frac{1-\cos Q_y}{2}\cos q_y\nonumber\\
& & +J_d(1-\cos Q_x\cos Q_y)\cos q_x\cos q_y, \nonumber
\end{eqnarray}
yielding the spin-wave dispersion $\omega_\bq=S\sqrt{A_\bq^2-B_\bq^2}$ and the inelastic structure factor at zero temperature\cite{Kruger+03}
\begin{equation}
\mathcal{S}_\textrm{inel}(\bq,\omega) = \sqrt{\frac{1-\gamma_\bq}{1+\gamma_\bq}}\delta(\omega-\omega_\bq)
\end{equation}
with $\gamma_\bq=B_\bq/A_\bq$. The resulting excitation spectra are shown in Fig.~\ref{swave} for different values of $\eta$. In the case of disordered orbitals, the $(\pi,0)$ antiferromagnet order is stable up to $\eta\approx 0.25$. Since $J_x=J_y$ the spectrum is gapless not only at the ordering wave vector $(\pi,0)$ but also at the antiferromagnetic wave vector $(\pi,\pi)$. However, the spectral weight is centered close to the ordering wave vector and goes strictly to zero at the antiferromagnetic wave vector. In the presence of orbital ordering the next-nearest neighbor couplings are anisotropic $J_x>J_y$ which in the case of the $(\pi,0)$-AF leads to a gap at the antiferromagnetic wave vector, $\Delta_{(\pi,\pi)}=2\sqrt{(2J_d-J_y)(J_x-J_y)}$. Since the anisotropy and the diagonal exchange are large in the orbital ferro state we find a very big gap at $(\pi,\pi)$. This gap reduces drastically for bigger $\eta$ where the orbital stripe state becomes favorable. Due to the large reduction of $J_d$ 
and also of the anisotropy, the gap is considerably smaller and continuously goes to zero as we approach the transition to the $(\pi,\pi)$-AF at $\eta\approx 0.14$ where  $2J_d-J_y=0$. This of course also leads to a strong anisotropy of the spin-wave velocities, $v_y/v_x=\sqrt{(2J_d-J_y)/(2J_d+J_x)}$.
On approaching the magnetic transition we find a significant softening of modes along the $(\pi,0)-(\pi,\pi)$ direction which leads to a considerable reduction of magnetic moments close to the transition.

\section{Discussion and Conclusions}

In the preceding we have derived and studied a spin-orbital Kugel-Khomskii Hamiltonian relevant to the Fe-As planes of the parent compound of the iron superconductors. A variety of interesting  spin and orbital ordered phases exist over a physical regime in parameter space. Due to the peculiarities of the pnictide lattice and this particular crystal field state we  show that the relevant Kugel-Khomskii model is of a particularly interesting kind. 

The essence of the 'spin-charge-orbital' physics is {\em dynamical frustration}. With so many 'wheels in the equation' it tends to be difficult to find solutions that satisfy simultaneously the desires of the various types of degrees of freedom in the problem. This principle underlies the quite complex phase diagrams of for instance manganites. But this dynamical frustration is also a generic property  of the spin-orbital models describing the Jahn-Teller degenerate Mott-insulators. In the 'classic' Kugel-Khomskii model\cite{KK} describing  $e_g$ degenerate $S=1/2$ $3d^9$ systems of cubic 3d systems, Feiner {\em et al.}\cite{feiner1,feiner2} discovered a point in parameter space where on the classical level this frustration becomes perfect. In the present context of pnictides this appears as particularly relevant since this opens up the possibility that quantum fluctuations can become quite important. 

We propose two specific orbital ordered phases that explain the orthorhombic transition observed in the experiments. These are orbital-ferro and orbital-stripe states. The orbital-stripe order is particularly interesting since it leads to a spin model that provides possible explanation for the reduction of magnetic moment. It is our main finding that in the idealized pnictide spin-orbital model the conditions appear optimal for the frustration physics to take over. We find large areas in parameter space where frustration is near perfect. The cause turns out to be a mix of intrinsic frustration associated with having $t_{2g}$ type orbital degeneracy, and the frustration of a geometrical origin coming from the pnictide lattice with its competing "$J1-J2$"  superexchange pathways. The significance of this finding is that  this generic frustration will render the spin-orbital degrees of freedom to be extremely soft, opening up the possibility that they turn into strongly fluctuating degrees of freedom -- a desired property when one considers pnictide physics.

We argued that the orthorhombic transition in half filled pnicitides and the associated anomalies in transport properties can be related to orbital order. When the parameters are tuned away from the frustration regime the main tendency of the system is to anti-ferro orbital ordering, which is the usual situation for antiferromagnets. An important result is that in the regime of relevance to the pnictides where the frustrations dominate we find phases that are at the same time $(\pi, 0)$ magnets and forms of orbital order that are compatible  with orthorhombic lattice distortions (Fig.'s 4,5). Besides the literal ferro-orbital ordered state (Fig. 5a), we find also a $(\pi, 0)$ or 'tweed' orbital order (Fig. 5b). This appears to be  the more  natural possibility in the insulating limit and if the weak superlattice reflections associated with this state would be observed this could be considered as a strong support for the literalness of the strong coupling limit.  Surely, the effects of itinerancy are expected to modify the picture substantially. Propagating fermions are expected to stabilize ferro-orbital orders~\cite{Kugel, kumarkampf}, which enhances the spatial anisotropy of the spin-spin interactions further~\cite{savrasov}.

Among the observable consequences of this orbital physics is its impact of the spin fluctuations.  We conclude the paper with an analysis of the spin waves in the orbital ordered phases, coming to the conclusion that also the spin sector is quite frustrated, indicating that the quantum spin fluctuations should be quite strong offering a rational for a strong reduction of the order parameter. 

Thus we have forwarded the hypothesis that the undoped iron pnictides are controlled by a very similar 'spin-charge-orbital' physics as found in ruthenates and manganites. To develop a more quantitative theoretical expectation is less straightforward and as it is certainly beyond standard LDA and LDA+U approaches will require investigations of correlated electron models such as we have derived here\cite{anisimovzaanenandersen,liechtensteinanisimovzaanen}, taking note of the fact that the pnictides most likely belong to the border line cases where the Hubbard $U$ is neither small nor large compared to the bandwidth\cite{sawatzkybrinkzaanen}.

\acknowledgments{The authors would like to acknowledge useful discussions with G. Giovannetti, J. Moore and G.A. Sawatzky. This work is financially supported by {\it Nanoned}, a nanotechnology programme of the Dutch Ministry of Economic Affairs and by the {\it Nederlandse Organisatie voor Wetenschappelijk Onderzoek~(NWO), and the Stichting voor Fundamenteel Onderzoek der Materie~(FOM)}.}

\appendix
\section{Effective Interaction Amplitudes}
By acting with the hopping operator $\Ham_t$ (\ref{Ht}) on all states in the ground state sector and calculating the overlap of the resulting states projected on the different intermediate states  we find the effective interaction amplitudes. For the high-spin intermediate state ($n=1$) we find by projecting on the intermediate $^4A_2$ multiplet,

\begin{eqnarray}
f_{zz}^{(1)} & = & \frac{4 t_{ab}^2-2(t_{aa}^2+t_{bb}^2)}{\epsilon(^4A_2)} \nonumber \\
f_{+-}^{(1)}& =& -\frac{4 t_{aa}t_{bb}}{\epsilon(^4A_2)}\nonumber \\
f_{++}^{(1)}& =& -\frac{4 t_{ab}^2}{\epsilon(^4A_2)}\nonumber \\
f_{zx}^{(1)}& =& \frac{4 t_{ab}(t_{bb}-t_{aa})}{\epsilon(^4A_2)}\nonumber \\
f_{z}^{(1)}& =& \frac{t_{bc}^2-t_{ac}^2}{\epsilon(^4A_2)+\Delta}\nonumber \\
f_{x}^{(1)}&= & -\frac{2 t_{ac}t_{bc}}{\epsilon(^4A_2)+\Delta}\nonumber \\
f_{0}^{(1)}& =&  \frac 12 \frac{2 t_{ab}^2+t_{aa}^2+t_{bb}^2}{\epsilon(^4A_2)}+ \frac{t_{ac}^2+t_{bc}^2}{\epsilon(^4A_2)+\Delta},
\end{eqnarray}
where the hopping matrix elements have to be specified for a particular bond. Likewise, we find by projections on the intermediate low-spin
states ($n=2$)

\begin{eqnarray}
f_{zz}^{(2)} & = & \frac 12 (2t_{ab}^2-(t_{aa}^2+t_{bb}^2))\nonumber\\
& & \times\left(\frac{4}{\epsilon(^2E)}-\frac{3}{\epsilon(^2T_1)}-\frac{3}{\epsilon(^2T_2)}\right)\nonumber \\
f_{+-}^{(2)}& =& \frac{2 t_{aa}t_{bb}}{\epsilon(^2E)}+3 t_{ab}^2\left(\frac{1}{\epsilon(^2T_1)}-\frac{1}{\epsilon(^2T_2)}\right) \nonumber \\
f_{++}^{(2)}& =& \frac{2 t_{ab}^2}{\epsilon(^2E)}+3t_{aa}t_{bb}\left(\frac{1}{\epsilon(^2T_1)}-\frac{1}{\epsilon(^2T_2)}\right) \nonumber \\
f_{zx}^{(2)}& =& t_{ab}(t_{bb}-t_{aa})\left( \frac{1}{\epsilon(^2E)}+\frac{3}{\epsilon(^2T_2)} \right)\nonumber \\
f_{z}^{(2)}& =& \frac 12 (t_{bc}^2-t_{ac}^2)\left(\frac{4}{\epsilon(^2E)+\Delta}+\frac{3}{\epsilon(^2T_1)+\Delta}\right.\nonumber \\
& & \left.+\frac{3}{\epsilon(^2T_2)+\Delta}-\frac{3(1-v_-^2)}{\epsilon(^2T_1^\Delta)}-\frac{3(1-v_+^2)}{\epsilon(^2T_2^\Delta)}\right)\nonumber \\
f_{x}^{(2)}&= & \frac 32 t_{ab}(t_{aa}+t_{bb})\left( \frac{1}{\epsilon(^2E)}+\frac{1}{\epsilon(^2T_1)} \right)\nonumber\\
& & +t_{ac}t_{bc} \left(\frac{2}{\epsilon(^2E)+\Delta}+\frac{3}{\epsilon(^2T_1)+\Delta}\right.\nonumber\\
& & \left.-\frac{3}{\epsilon(^2T_2)+\Delta}+\frac{3(1-v_-^2)}{\epsilon(^2T_1^\Delta)}+\frac{3(1-v_+^2)}{\epsilon(^2T_2^\Delta)}\right)\nonumber \\
f_{0}^{(2)}& =&  \frac 18 (2t_{ab}^2+t_{aa}^2+t_{bb}^2)\nonumber\\
& & \times\left(\frac{4}{\epsilon(^2E)}+\frac{3}{\epsilon(^2T_1)}+\frac{3}{\epsilon(^2T_2)}\right)\nonumber\\
& & +\frac 12 (t_{ac}^2+t_{bc}^2)\left(\frac{4}{\epsilon(^2E)+\Delta}+\frac{3}{\epsilon(^2T_1)+\Delta}\right.\nonumber\\
& & \left.+\frac{3}{\epsilon(^2T_2)+\Delta}+\frac{3(1-v_-^2)}{\epsilon(^2T_1^\Delta)}+\frac{3(1-v_+^2)}{\epsilon(^2T_2^\Delta)}\right)\nonumber\\
& & +3 t_{cc}^2\left(\frac{1-v_-^2}{\epsilon(^2T_1^\Delta)+\Delta}+\frac{1-v_+^2}{\epsilon(^2T_2^\Delta)+\Delta}  \right).
\end{eqnarray}

The terms bilinear in the pseudospin operators result solely from hopping processes involving the $|a\rangle$ and $|b\rangle$ orbitals only. The hoppings between
the $|c\rangle := |xy\rangle$ orbitals enter only as a positive constant in $\mathcal{Q}^{(2)}$ leading to a conventional antiferromagnetic superexchange contribution. Interestingly, the coupling between the $|c\rangle$ and $|a\rangle$,$|b\rangle$ orbitals results in magnetic field terms for the orbital pseudospins. 

\section{Hopping Matrix Elements}

For a given As-Fe bond with direction cosines $l,m,n$, the $p$ to $t_{2g}$ hoppings are given by
\cite{slaterkoster}

\begin{eqnarray}
t_{x,zx} & = & n[\sqrt{3}l^2 (pd\sigma) + (1-2l^2)(pd\pi)] \nonumber \\
t_{x,yz} & = & lmn[\sqrt{3} (pd\sigma) - 2(pd\pi)]  \nonumber \\
t_{x,xy} & = & m[\sqrt{3}l^2 (pd\sigma) + (1-2l^2)(pd\pi)] \nonumber \\
t_{y,zx} & = & t_{x,yz}  = t_{z,xy} \nonumber \\
t_{y,yz} & = & n[\sqrt{3}m^2 (pd\sigma) + (1-2m^2)(pd\pi)] \nonumber \\
t_{y,xy} & = & l[\sqrt{3}m^2 (pd\sigma) + (1-2m^2)(pd\pi)] \nonumber \\
t_{z,zx} & = & l[\sqrt{3}n^2 (pd\sigma) + (1-2n^2)(pd\pi)] \nonumber \\
t_{z,yz} & = & m[\sqrt{3}n^2 (pd\sigma) + (1-2n^2)(pd\pi)].  
\end{eqnarray}

Using direction cosines $(l,m,n)$ ($l^2+m^2+n^2=1$) with $|l|=|m|$ resulting from the orthorhombic symmetry, we find that only the following hopping-matrix elements are non-zero,

\begin{eqnarray}
t_{aa}^x = t_{bb}^y & =:& t_1,\nonumber \\ 
t_{bb}^x =  t_{aa}^y & =: & t_2\nonumber \\
t_{aa}^d = t_{bb}^d & =: & t_3\nonumber \\
t_{ab}^{d-} = -t_{ab}^{d+} & =: & t_4\nonumber \\
t_{cc}^x = t_{cc}^y & =: & t_5\nonumber \\
t_{cc}^d & =: & t_6\nonumber \\
t_{ac}^x=t_{bc}^y & =: & t_7.
\end{eqnarray}

These hopping matrix elements which are shown schematically in Fig.~\ref{Fe_hopping} can be parametrized by the lattice parameter $\lambda=|n/l|$ and the ratio $\gamma=(pd\pi)/(pd\sigma)$ as

\begin{eqnarray}
t_1/t & = & -2(B^2-A^2-C^2) \nonumber \\
t_2/t & = & -2(B^2-A^2+C^2) \nonumber \\
t_3/t & = & -(B^2+A^2-C^2) \nonumber \\
t_4/t & = & 2AB - C^2 \nonumber \\
t_5/t & = & 2 A^2 \nonumber \\
t_6/t & = & 2(B/\lambda)^2-A^2 \nonumber \\
t_7/t & = & 2(AC+AB/\lambda-B^2/\lambda),
\end{eqnarray}
where we have introduced the overall energy scale $t=(pd\sigma)^2/\Delta_{pd}$ and defined for abbreviation

\begin{eqnarray}
A & = & \frac{\lambda(\sqrt{3}-2\gamma)}{\sqrt{2+\lambda^2}^3} \nonumber \\
B & = & \frac{\lambda(\sqrt{3}+\lambda^2\gamma)}{\sqrt{2+\lambda^2}^3}\nonumber  \\
C & = & \frac{\sqrt{3}\lambda^2+(2-\lambda^2)\gamma}{\sqrt{2+\lambda^2}^3}.
\end{eqnarray}

\section{Orbital Part of the Hamiltonian}

For the nearest neighbor bonds along $\hat{x}$ and $\hat{y}$ the orbital operators in the spin-orbital Hamiltonian are given by  

\begin{eqnarray}
\hat{\Omega}_{x,y} & = & \frac 12 (\alpha_1^2+\alpha_2^2)(1+2\eta r_1-\eta r_3)\hT_i^z\hT_j^z\nonumber\\
& & +\alpha_1\alpha_2 (1+2\eta r_1+\eta r_3)\hT_i^x\hT_j^x\nonumber\\
& & +\alpha_1\alpha_2 (1+2\eta r_1-\eta r_3)\hT_i^y\hT_j^y\nonumber\\
& & \mp\frac{1}{12}\alpha_7^2 (7\tilde{r}_2+3\tilde{r}_3-2\tilde{r}_1-3g_1)(\hT_i^z+\hT_j^z)\nonumber\\
& & +\frac 18 (\alpha_1^2+\alpha_2^2)(1-2\eta r_1-\eta r_3)\nonumber\\
& & +\frac{1}{12}\alpha_7^2 (7\tilde{r}_2+3\tilde{r}_3-2\tilde{r}_1+3g_1)+\frac 12 \alpha_5^2 g_2\\
\hat{\Gamma}_{x,y} & = & \frac 12 (\alpha_1^2+\alpha_2^2)\eta( r_1+r_3)\hT_i^z\hT_j^z\nonumber\\
& & +\alpha_1\alpha_2\eta (r_1- r_3)\hT_i^x\hT_j^x+\alpha_1\alpha_2\eta (r_1+r_3)\hT_i^y\hT_j^y\nonumber\\
& & -\frac 18 (\alpha_1^2+\alpha_2^2) (2+\eta r_1-\eta r_3)\nonumber\\
& & -\frac{1}{12}\alpha_7^2 (\tilde{r}_1+7\tilde{r}_2+3\tilde{r}_3+3g_1)-\frac 12 \alpha_5^2 g_2,
\end{eqnarray}
where we have defined $\tilde{r}_1=  1/(1-3\eta+\delta)$, $\tilde{r}_2=\tilde{r}_1|_{\eta=0}$, $\tilde{r}_3=  1/(1+2\eta+\delta)$,
and $r_i=\tilde{r}_i|_{\delta=0}$ and introduced the functions 

\begin{eqnarray}
g_1 & = &  \frac{1+\eta+\delta}{1+2\eta-\delta^2}\\
g_2 & = &  \frac{1+\eta+2\delta}{1+2\eta(1+\delta)+2\delta}.
\end{eqnarray}

Likewise, for the bonds along the $\hat{x}\pm\hat{y}$ diagonals we obtain

\begin{eqnarray}
\hat{\Omega}_{d\pm} & = & (\alpha_3^2-\alpha_4^2)(1+2\eta r_1-\eta r_3)\hT_i^z\hT_j^z\nonumber\\
& & + (\alpha_3^2+\alpha_4^2)(1+2\eta r_1+\eta r_3)\hT_i^x\hT_j^x\nonumber\\
& & + (\alpha_3^2-\alpha_4^2)(1+2\eta r_1-\eta r_3)\hT_i^y\hT_j^y\nonumber\\
& & \mp \alpha_3\alpha_4 (\hT_i^x+\hT_j^x)\nonumber\\
& & +\frac 14 (\alpha_3^2+\alpha_4^2) (1-2\eta r_1-\eta r_3)\nonumber\\
& & +\frac 12 \alpha_6^2 g_2\\
\hat{\Gamma}_{d\pm} & = & (\alpha_3^2-\alpha_4^2)\eta (r_1+r_3)\hT_i^z\hT_j^z\nonumber\\
& & + (\alpha_3^2+\alpha_4^2)\eta (r_1-r_3)\hT_i^x\hT_j^x\nonumber\\
& & + (\alpha_3^2-\alpha_4^2)\eta (r_1+r_3)\hT_i^y\hT_j^y\nonumber\\
& & -\frac 14 (\alpha_3^2+\alpha_4^2) (2+\eta r_1-\eta r_3)\nonumber\\
& & -\frac 12 \alpha_6^2 g_2.
\end{eqnarray}

\bibliography{strings,refs}
\bibliographystyle{apsrev}

\end{document}